# Four kinetic energy formulas for studying the interaction of energetic ions with carbon nanotubes

Li-Ping Zheng, Jian-Qing Cao, Can Huang, Guo- Liang Ma

Shanghai Institute of Applied Physics, Chinese Academy of Sciences, P.O. Box 800-204, Shanghai 201800, China

**Abstract** Three questions should be interesting, for better understanding of the basic physics of damage creation and stopping ions of different types, in the damaged tube: 1) Between the mass and charge effects, i.e. potential- independent and potential-dependent effects on the damaged tube, Which ones may the cross- section $\sigma$ effects belong to? 2) One method has calculated number of energetic recoils (both ions and carbon atoms), which hit the metal layer below the damaged tube, to study the stopping of both incident ions and recoil carbon atoms in it. Another method has calculated the ion- deposited energy $E_0 - E_{penetrating}$ in the damaged tube, to distinguish between penetrating and stopped ions. Which method is better for studying the stopping between two ones? 3) Why can tens of the graphitic shells easily stop energetic ions with energies up to $10\,keV$? Which effects on the stopping? In kinetic energy formulas, we give answers for discussion.

# 1. Introduction

The interaction of energetic ions with carbon nanotubes plays a fundamental role both in the field of ion beam irradiation and in nanoscience. The rising interest in irradiation effects in carbon nanostructures is triggered, for instance, by recent observations of intriguing irradiation- induced phenomena [1-4]. To understand the basic physics of damage creation and stopping ions of different types, it is instructive to dwell first upon the irradiation of SWCNTs (single-walled carbon nanotubes) due to their simple and well- defined structure, between SWCNTs and MWCNTs (multi-walled carbon nanotubes) [5].

Krasheninnikov et al. have classical molecular dynamics (MD) simulation works with the ZBL potential potentials, such as a) 'Carbon nanotubes as masks against ion irradiation' [1], b) 'Ion ranges and irradiation- induced defects in multi- walled carbon nanotubes' [2, 3] and c) 'Relative abundance of single and double vacancies in irradiated single- walled carbon nanotubes' [4] etc. Their works [1-4] might originate from the works of 'Stopping of energetic ions in carbon nanotubes' [5]. Of course, in their works [1-5], nuclear stopping dominates over electronic stopping [1-9]. As discussed in section 2, in our opinion, three questions should be interesting, for better understanding of the basic physics of damage creation and stopping ions of different types, in the damaged tube.

According to classical theories [6, 7], energetic ions and atoms can be seen as particles with nuclear point charge and nuclear point mass. In particle collisions, the well- known $4Mm/(M+m)^2$ ratio is based on the conservation of energy and that of momentum. However, the $p$ impact parameter and then the $\theta$ ion- scattered angle necessarily exist in the interaction potentials of nuclear point charges [9]. Therefore, the $4Mm/(M+m)^2$ effects belong to nuclear point mass ones, i.e. potential- independent effects, but the $p$ impact parameter effects and the $\theta$ ion- scattered angle ones belong to nuclear point charge effects, i.e. potential- dependent ones. Here, $M$ and $m$ are representative of the energetic ion mass and the carbon atom one, respectively. Detail analysis is in section 2.

## 2. Kinetic energy formula and discussion

**The question one**: Krasheninnikov et al. have studied the interaction of energetic ions with carbon nanotubes up to $keV$ ion- irradiation energies. For damage studies, they found the irradiation- induced damage which proved to be higher for heavy ions than for light ones due to higher values of the cross- section $\sigma$ for the defect production in the SWCNT [5]. Here are either the mass or the charge effects, i.e. potential- independent and potential- dependent effects on the SWCNT. Which ones may the cross- section $\sigma$ effects belong to?

**The question two:** Krasheninnikov et al. have calculated number of energetic recoils (both ions and carbon atoms) [1, 5], which hit the metal layer below the SWCNT, to study the stopping of both incident ions and recoil carbon atoms in it (Fig. 1 of Ref. [1]). If this number equals 0, both incident ions and recoil carbon atoms stop in the SWCNT. In our opinion, their method of calculating stopping means enough.

If the incident energy $E_0$ is high enough, the incident ion easily penetrates through the SWCNT or the MWCNT while it becomes the penetrating ion with the $E_{penetrating}$ kinetic energy, so $E_0 - E_{penetrating}$ equals the ion- deposited energy in the SWCNT or the MWCNT. We have calculated the ion- deposited energy $E_0 - E_{penetrating}$ to distinguish between penetrating and stopped ions [9]. Obviously, if kinetic energy $E_{penetrating} > 0$, i.e. $E_0 > E_0 - E_{penetrating}$ formula (1), the incident ion becomes the penetrating one; if kinetic energy $E_{penetrating} = 0$, i.e. $E_0 = E_0 - E_{penetrating}$ formula (2), the incident ion becomes the stopped one. In our opinion, in Ref. [9], our method of calculating stopping means necessary. Which methods are better for studying nuclear stopping between [1, 5] and [9]?

**The question three:** At $0.3 keV < E_0 \leq 1 keV$ incident energy, Krasheninnikov et al. have found that ions lose about 0.3 $keV$ of their kinetic energy in a collision with one shell, i.e. the SWCNT [5], so they have given that the MWCNT usually has tens of the graphitic shells, it can easily stop energetic ions with energies up to 10 $keV$ and thus can be used for spatially selective ion implantation. Why can the MWCNT with tens of the graphitic shells easily stop energetic ions with energies up to $10\,keV$? Which effects on the stopping?

**The answer one:** According to classical theories [6, 7], energetic ions and atoms can be seen as particles with nuclear point charge and nuclear point mass. In particle collisions, the well-known $4Mm/(M+m)^2$ ratio is based on the conservation of energy and that of momentum. However, the $p$ impact parameter and then the $\theta$ ion-scattered angle necessarily exist in the interaction potentials of nuclear point charges. Therefore, the $4Mm/(M+m)^2$ effects belong to nuclear point mass ones, i.e. potential-independent effects, but the $p$ impact parameter effects and the $\theta$ ion-scattered angle ones belong to nuclear point charge effects, i.e. potential-dependent ones. Evidently, the cross-section $\sigma$ effects ($\sigma = \pi p^2$) belong to the nuclear point charge ones.

**The answer three:** Under various energetic ion irradiations, at the same incident energy, i.e. this $E_0$ kinetic energy keeps constant, in a binary collision model based on the Moliere potential for the ion-atom (or atom-atom) interaction, we list an ion-deposited energy formula (formula 7 of Ref. [9]) for the irradiated MWCNT (or SWCNT) here, that is

$$E_0 - E_{penetrating} = (4Mm/(M+m)^2)\sum_{i=0}^{n-1} E_i \sin^2(\theta_{i+1}/2) \qquad (3)$$

Where, the final ion-kinetic-energy $E_{penetrating}$ in $n$ collisions; the $E_0 - E_{penetrating}$ ion-deposited energy equals the product of the $4Mm/(M+m)^2$ ratio and the $\sum_{i=0}^{n-1} E_i \sin^2(\theta_{i+1}/2)$ angle-correlated energy. Evidently, this formula studies relation between the nuclear point charge effects and its mass ones on the $E_0 - E_{penetrating}$ ion-deposited energy.

As mentioned above, if kinetic energy $E_{penetrating} = 0$, i.e. $E_0 = E_0 - E_{penetrating}$ formula (2), the incident ion becomes the stopped one in the MWCNT (or SWCNT). We combine formula (2) and formula (3), that is

$$E_0 = E_0 - E_{penetrating} = (4Mm/(M+m)^2)\sum_{i=0}^{n-1} E_i \sin^2(\theta_{i+1}/2) \qquad (4)$$

In formula (4), formula $E_0 = (4Mm/(M+m)^2)\sum_{i=0}^{n-1} E_i \sin^2(\theta_{i+1}/2)$ indicates that the $4Mm/(M+m)^2$ effects and the $\sum_{i=0}^{n-1} E_i \sin^2(\theta_{i+1}/2)$ angle- correlated energy effects cancel each other, while the incident energy $E_0$ is conservative, i.e. it keeps constant. Correspondently, the formula (2) $E_0 = E_0 - E_{penetrating}$ indicates that the incident ion becomes the stopped one. In short, because two nuclear point effects cancel each other, the MWCNT with tens of the graphitic shells can easily stop energetic ions with energies up to $10\,keV$.

We have proposed [9] that three nuclear point effects, i.e., the $4Mm/(M+m)^2$ effects, the $\sum_{i=0}^{n-1} E_i \sin^2(\theta_{i+1}/2)$ angle- correlated energy ones, and the cross- section $\sigma$ ones on tube- damage and stopped ions of different types. As shown in Fig. 3 of Ref. [5]，under $keV$- noble ion irradiations of 1 shell, the cross- section $\sigma$ for the defect production increases from light to heavy ions, as the ion- nuclear point charge $Ze$ increases [9] ($Z$ is atomic number of the ion, and $e$ is the charge on a proton.). Correspondently, the coordination defect number increases from light to heavy noble ions (Fig. 2 of Ref. [5]). Namely, the cross section $\sigma$ effects, i.e. nuclear point charge ones, dominate the tube damage. This is our further answer of the question one.

**3. Addition**

As shown in Fig. 4 of Ref. [9], under the same hundreds $keV$- noble ion irradiations of 1 shell, the coordination defect number maximizes, while its mass has an intermediate mass ($^{20}$Ne) value. This is because the $4Mm/(M+m)^2$ effects, i.e. the nuclear point mass ones, dominate over the $\sum_{i=0}^{n-1} E_i \sin^2(\theta_{i+1}/2)$ effects, i.e. nuclear point charge ones, on the tube- damage. Note that a Ne ion has the most $4Mm/(M+m)^2$ value in He, Ne, Ar, Kr and Xe ions. In $E_0 > E_0 - E_{penetrating}$ formula (1), Ref. [5, 9] has shown two different damage phenomena in the SWCNT, but in $E_0 = E_0 - E_{penetrating}$ formula (2), would be the third damage phenomena found in the MWCNT?

## 4. Summary

For better understanding of the basic physics of damage creation and stopping ions of different types, in the damaged tube, we answer three questions as followers: **a)** the cross- section $\sigma$ ($\sigma = \pi p^2$) effects for the defect production belong to the nuclear point charge ones. **b)** If kinetic energy $E_{penetrating} > 0$, i.e. $E_0 > E_0 - E_{penetrating}$ formula (1), the incident ion becomes the penetrating one; if kinetic energy $E_{penetrating} = 0$, i.e. $E_0 = E_0 - E_{penetrating}$ formula (2), the incident ion becomes the stopped one. **C)** Because two nuclear point effects (the $4Mm/(M+m)^2$ effects and the $\sum_{i=0}^{n-1} E_i \sin^2(\theta_{i+1}/2)$ angle- correlated energy effects) cancel each other, the MWCNT with tens of the graphitic shells can easily stop energetic ions with energies up to $10\,keV$.

**Acknowledgements**

Supported by National Basic Research Program of China (973 Program) 2010CB832903, National Natural Science Foundation of China No 11175232 and No 11175235.

**References**

[1] A.V. Krasheninnikov, K. Nordlund, J. Keincnen, Appl. Phys. Lett. **81 (2002)**

1101.

[2] A.V. Krasheninnikov, K. Nordlund, J. Appl. Phys. **107 (2010)** 071301.

[3] J. Pomoell, A.V. Krasheninnikov, K. Nordlund, J. Keincnen, J. Appl. Phys. **96 (2004)** 2864.

[4] A. Tolvanen, J. Kotakoski, A. V. Krasheninnikov, K. Nordlund, Appl. Phys. Lett. **91 (2007)** 173109.

[5] J. Pomoell, A.V. Krasheninnikov, K. Nordlund, J. Keincnen, Nucl. Instrum. Methods **2003**, B 206, 18.

[6] P. Sigmund, << Particle Penetration and Radiation Effects: General Aspects and Stopping of Point Charge>>, Springer，2006.

[7] P. Sigmund, Phys. Rev. **184 (1969)** 383.

[8] Biersack, J. P.; Haggmark, L. G. Nucl. Instrum. Methods **1980**, 174, 257.

[9] Li-Ping Zheng, Long Yan, Zhi-Yong Zhu, Guo-Liang Ma. Appl. Phys. A (**2016**) 122:222.